# Non-iterative Optimization Algorithm for Active Distribution Grids Considering Uncertainty of Feeder Parameters

Jiexuan Wu[1], Mingbo Liu[1], Wentian Lu[1]*, Kaijun Xie[1], Min Xie[1]
[1]School of Electric Power Engineering, South China University of Technology, Guangzhou 510640, China

***Abstract*—To cope with fast-fluctuating distributed energy resources (DERs) and uncontrolled loads, this paper formulates a time-varying optimization problem for distribution grids with DERs and develops a novel non-iterative algorithm to track the optimal solutions. Different from existing methods, the proposed approach does not require iterations during the sampling interval. It only needs to perform a single one-step calculation at each interval to obtain the evolution of the optimal trajectory, which demonstrates fast calculation and online-tracking capability with an asymptotically vanishing error. Specifically, the designed approach contains two terms: a prediction term tracking the change in the optimal solution based on the time-varying nature of system power, and a correction term pushing the solution toward the optimum based on Newton's method. Moreover, the proposed algorithm can be applied in the absence of an accurate network model by leveraging voltage measurements to identify the true voltage sensitivity parameters. Simulations for an illustrative distribution network are provided to validate the approach.**

***Index Terms*—Non-iterative algorithm, time-varying optimization, distributed energy resources (DERs), prediction-correction (PC) method, uncertain feeder parameters.**

## NOMENCLATURE

*A. Indices and Sets*

| | |
|---|---|
| $t$ | Index for time |
| $t_k, k$ | Index for time instant |
| $i$ | Index for nodes |
| $l$ | Index for lines |
| $N$ | Set of nodes |

*B. Parameters*

| | |
|---|---|
| $\mathbf{V}_{nom}$ | Voltage nominal value |
| $\boldsymbol{u}^{\min} / \boldsymbol{u}^{\max}$ | Lower/upper bounds of control variables |
| $\boldsymbol{P}_L / \boldsymbol{Q}_L$ | Real/reactive power loads |
| $\boldsymbol{A}, \boldsymbol{B}$ | Voltage sensitivity matrixes |
| $\boldsymbol{M}$ | Node-to-branch incidence matrix |
| $\varpi_i$ | Feasible operating range of $P_{gi}$ and $Q_{gi}$ |
| $C_{pi} / C_{qi}$ | Cost coefficients of DERs' real/reactive power |
| $P_i^{av} / Q_i^{av}$ | Available real/reactive power of DER $i$ |
| $\theta_i$ | Power factor angle of DER $i$ |
| $c$ | Barrier parameter |
| $s$ | Slack function |

*C. Variables*

| | |
|---|---|
| $\boldsymbol{P}_g / \boldsymbol{Q}_g$ | Output real/reactive power vectors |
| $\boldsymbol{P} / \boldsymbol{Q}$ | Net real/reactive power vectors |
| $\boldsymbol{V}$ | Voltage magnitude vector |
| $\boldsymbol{u}$ | Control variable vector |
| $P_{gi} / Q_{gi}$ | Real/reactive power outputs of DER $i$ |
| $V_i$ | Voltage magnitude at node $i$ |

## I. INTRODUCTION

DISTRIBUTED energy resources (DERs) are envisioned to be dispersed in future distribution networks through power electronic devices to reduce carbon footprints [1]. Under the influence of ambient conditions, some renewable energy sources, such as photovoltaic (PV) systems and wind turbines (WTs), show fast time-varying characteristics. For example, the output of PV may change dramatically within a few seconds when affected by clouds, easily causing network security and voltage quality problems [2]. Additionally, DC/AC converter and electronic modulation technologies are widely used in distributed power supplies [3], which enables them to be rapidly regulated, e.g., in 0.02 s (for 50 Hz frequency). This feature suggests that DERs' fast-responding characteristics can enhance the active operation and regulation ability of distribution networks if the optimal determination of DERs' setpoints can match their adjustment time.

The main goal of time-varying optimization methods is to track the optimal trajectories of continuously varying optimization problems (within the allowable error range) [4]. A natural approach is to sample the problems at specific times and then solve the resulting sequence of time-invariant optimization problems using iterative algorithms [5]. Offline algorithms [6, 7] are used to solve problems that change slowly over time. A dual $\varepsilon$-subgradient method was proposed in [8] to seek inverter setpoints of PVs to bridge the temporal gap between long-term system optimization and real-time inverter control. In fast-changing settings, measurement-based online methods [2], [9]–[13] have been developed on a timescale of seconds. A feedback control strategy for optimal reactive power setpoints for microgenerators was proposed in [10] to provide real-time reactive power compensation. Online primal-dual-type methods were applied to develop real-time feedback algorithmic frameworks for time-varying optimal power flow

* Corresponding author.
E-mail:1324536158@qq.com(Jiexuan Wu); epmbliu@scut.edu.cn(Mingbo Liu); hnlgtiantian@163.com(Wentian Lu); ep_xiekj@mail.scut.edu.cn(Kaijun Xie); minxie@scut.edu.cn(Min Xie).

problems in [11]–[13].

The critical feature shared by the above-mentioned methods is that they only utilize the current information of the problem parameters. In other words, these algorithms do not perform a "prediction" step; rather, they only carry out "correction" steps once the current information is obtained [4]. These approaches need iterations during the sampling interval to converge toward the optimum of the sampled time-invariant problem, while the actual solution drifts with time. Therefore, these approaches are likely to induce a large tracking error [14]. To reduce tracking error, discrete-time prediction-correction (PC) methods were proposed in [15]–[21] utilizing the prediction information of the problem parameters to identify the change of the optimum and correct the prediction based on the newly acquired information. Specifically, Simonetto *et al*. [21] proposed a discrete-time PC approach to implement time-varying convex optimization and extended it to DERs operation optimization in distribution systems, while other algorithms in [15]–[20] are temporarily applied to simple mathematical examples. However, the discrete PC methods still cannot fully identify the dynamics within the interval, whose asymptotical error depends on the length of sampling interval and iteration steps of "correction." To track the optimal solution with an asymptotically vanishing error, a continuous-time PC method was designed for an unconstrained optimization problem [22]. In [14], the authors developed a PC interior-point method to solve constrained time-varying convex optimization problems and implemented it in a sparsity promoting least squares problem and a collision-free robot navigation problem. In this paper, we will extend this method to develop a non-iterative PC algorithm to identify the optimal power setpoints of DERs and network voltages.

The exact relationship between nodal voltages and power injections in distribution networks is nonlinear, which will cause the optimization model to be non-convex and difficult to solve. A pivotal approach in most online methods is to approximate the nonlinear power flow by a linear model, e.g., in [2], [10]–[13], [21]. However, these schemes assume that we can obtain an ideal model of the distribution network, so they may not work properly when an accurate model is not available or feeder parameters change. For this situation, data-driven methods [23]–[26] provide some good ideas. A least squares estimator was utilized in [23] to compute voltage magnitude and power loss sensitivity coefficients in a low voltage network. Forward and inverse regression methods [24] and a recursive weighted least squares method [25] were studied for the case in which the model of the distribution system is not completely known. Different from the approaches mentioned above that estimate all the sensitivity elements (requiring a great deal of measurements and a high sampling rate), the authors in [26] proposed an approach to reduce the number of parameters to be estimated exploiting the structural characteristics of balanced radial distribution networks. In this paper, we involve a similar strategy for distribution networks, in which voltage measurements throughout the feeder are collected to identify the true voltage sensitivity parameters based on a linear power flow model.

To better track the dynamics within the interval, a non-iterative PC algorithm for distribution grids is developed in this paper to identify the time-varying optimal power setpoints of DERs with an asymptotically vanishing error. The fast-changing nature of renewable energy resources and uncontrolled loads is taken into account in the prediction term in order to identify the change of the optimum; a correction term represented by a Newton's method pushes the solution toward the optimum. Moreover, in the presence of uncertainty in the feeder parameters or the absence of an exact model of the system, voltage sensitivity parameters are identified by measurement information. By doing so, the optimal trajectories of DERs' power setpoints and network voltages will automatically adapt to system disturbance, such as renewable energy sources, uncontrollable loads, and system model parameters.

The main contributions of this paper follow.

1) A non-iterative algorithm with "prediction" and "correction" terms is constructed to track the optimal solutions of distribution grids' time-varying optimization with an asymptotically vanishing error. Compared with the existing methods, the proposed method is considerably faster because of its non-iterative nature, which makes the algorithm suitable to cope with the rapid changes of renewable energy resources. For example, the proposed algorithm can achieve fast calculation, so as to match the adjustment time of the inverter (i.e., 0.02 s). Moreover, the designed algorithm, based on "prediction" and "correction" terms, can obtain the evolution of the optimal trajectory, which demonstrates online-tracking capability with an asymptotically vanishing error.

2) Exploiting the structural characteristics of distribution networks, voltage sensitivity matrices are obtained based on a few online measurements, being different from methods of offline estimation followed by online application. This makes the designed approach very well suited for applications in the absence of an accurate network or with variational feeder parameters while without incurring extra computational burden.

The remainder of this paper is organized as follows. Section II formulates the time-varying optimization problem of distribution grids with DERs. In Section III, a non-iterative PC algorithm based on voltage measurements is proposed. Section IV presents simulation results on an illustrative system. The conclusion is provided in Section V.

## II. Problem Formulation

Consider a distribution feeder composed of $n$ + 1 nodes collected in the set $N \cup \{0\}$ with $N$: = {1, …, $n$} and $L$ distribution lines. Node 0 represents the point of common coupling (PCC) or substation. A time-varying minimization problem that captures the optimization objective, operation constraints, and the power flow equations is formulated as (1), which enables us to capture the variability of ambient conditions and noncontrollable energy assets.

$$\min_{\boldsymbol{P}_g,\boldsymbol{Q}_g} f\left(P_{gi},Q_{gi},t\right)=\sum_{i\in N} C_i\left(P_{gi},Q_{gi},t\right)+\frac{\gamma}{2}\left\|\boldsymbol{V}-\mathbf{V}_{nom}\right\|^2, \quad (1a)$$

$$s.t. \quad \left(P_{gi},Q_{gi}\right)\in\varpi_i\left(t\right),\forall i\in N, \quad (1b)$$

$$V_i^{\min}\le V_i\left(\boldsymbol{P}\left(t\right),\boldsymbol{Q}\left(t\right)\right)\le V_i^{\max},\forall i\in N, \quad (1c)$$

where $C_i(P_{gi}, Q_{gi}, t) = C_{pi}(P_{gi} - P_i^{tar}(t))^2 + C_{qi} Q_{gi}^2$ represents the cost objective associated with DERs; $C_{pi}$ and $C_{qi}$ are cost coefficients of real and reactive power of DER $i$, respectively; $P_{gi}$ and $Q_{gi}$ are real and reactive power setpoints of DER $i$, respectively; $\boldsymbol{P}_g = [P_{g1}; P_{g2};\ldots; P_{gn}]$ and $\boldsymbol{Q}_g = [Q_{g1}; Q_{g2};\ldots; Q_{gn}]$; $\gamma$ is a positive constant; $\varpi_i(t)$ represents the feasible operating range of $P_{gi}$ and $Q_{gi}$. $V_i$ denotes voltage magnitude at node $i$, and $\boldsymbol{V} = [V_1;V_2;\ldots;V_n]$; $\mathbf{V}_{nom}$ denotes the voltage nominal value. The function $V_i(\boldsymbol{P}(t), \boldsymbol{Q}(t))$ captures the relationship between voltage magnitudes and net real and reactive power vectors $\boldsymbol{P}(t)$, $\boldsymbol{Q}(t)$.

Specifically, for PV systems and WTs, $P_i^{tar}(t) = P_i^{av}(t)$ in (1a), with $P_i^{av}(t)$ denoting the maximum available real power of PV or WT at node $i$. The set $\varpi_i$ is given by:

$$\varpi_i(t) = \left\{ \begin{array}{l} 0 \le P_{gi} \le P_i^{av}(t), \left|Q_{gi}\right| \le Q_i^{av}(t), \\ Q_i^{av}(t) = P_i^{av}(t)\tan\theta_i \end{array} \right\},$$

where $Q_i^{av}(t)$ represents the reactive power limitation of DER $i$; $\theta_i$ is the power factor angle.

As for energy storage systems (ESSs), $P_i^{tar}(t) = 0$ in (1a), and the set $\varpi_i$ is given by:

$$\varpi_i(t) = \left\{ \begin{array}{l} -P_{i,\max}^{ch} \le P_{gi} \le P_{i,\max}^{dis}, \\ -\frac{1}{\eta_c}\left(W_{i,\max} - W_{i,0}\right) \le P_{gi} \le \eta_d\left(W_{i,0} - W_{i,\min}\right) \end{array} \right\},$$

where $P_{i,\max}^{ch}$ / $P_{i,\max}^{dis}$ denote the rated charging/discharging power of the ESS at node $i$; $\eta_c / \eta_d$ are charging/discharging efficiency; $W_{i,\min} / W_{i,\max}$ denote the lower/upper bound of the total energy storage of an ESS. $W_{i,0}$ represents the current energy storage of an ESS, which can be calculated from the information at the previous moment.

Let $\boldsymbol{P}(t) = \boldsymbol{P}_g + \boldsymbol{P}_L(t)$ and $\boldsymbol{Q}(t) = \boldsymbol{Q}_g + \boldsymbol{Q}_L(t)$. The nonlinear relationship between $\boldsymbol{V}$, $\boldsymbol{P}$, and $\boldsymbol{Q}$ can be approximated by a linear model [16] as follows:

$$\boldsymbol{V} = \boldsymbol{A}\boldsymbol{P}(t) + \boldsymbol{B}\boldsymbol{Q}(t) + \mathbf{c}, \tag{2}$$

where $\mathbf{c} = \mathbf{1}_n$ is an $n$-dimensional all-ones vector; parameters $\boldsymbol{A} \subset \mathbb{R}^{n\times n}$, $\boldsymbol{B} \subset \mathbb{R}^{n\times n}$, which are called voltage sensitivities (see, e.g., [26]), are time varying and can be estimated from a few measurements using an effective data-driven algorithm in this paper, and we will describe in detail in Section III how to estimate these time-varying parameters.

Let $\boldsymbol{u} = [\boldsymbol{P}_g; \boldsymbol{Q}_g]$ be the control variables of problem (1), $\boldsymbol{D} = [\boldsymbol{A}\ \boldsymbol{B}]$, and $\boldsymbol{W}(t) = [\boldsymbol{P}_L(t); \boldsymbol{Q}_L(t)]$ capture the noncontrollable load. Then the time-varying optimization problem (1) can be rearranged as following quadratic form:

$$\min_{\boldsymbol{u}} \quad f(\boldsymbol{u},t) = \boldsymbol{u}^T \frac{\boldsymbol{K}}{2}\boldsymbol{u} + \boldsymbol{d}(t)^T \boldsymbol{u} + \frac{\gamma}{2}\left\|\boldsymbol{V} - \mathbf{V}_{nom}\right\|^2, \tag{3a}$$

$$s.t. \quad \boldsymbol{u}^{\min}(t) \le \boldsymbol{u} \le \boldsymbol{u}^{\max}(t), \tag{3b}$$

$$\boldsymbol{V}^{\min} \le \boldsymbol{V} = \mathbf{c} + \boldsymbol{D}\boldsymbol{u} + \boldsymbol{D}\boldsymbol{W}(t) \le \boldsymbol{V}^{\max}, \tag{3c}$$

where $\boldsymbol{K} = diag(2C_{pi}, 2C_{qi}, i \in N)$; $\boldsymbol{d}(t) = [-2C_{pi} P_i^{tar}(t), \mathbf{0}]^T$, $i \in N$ and $\mathbf{0}$ is a $(1 \times n)$-dimensional vector, all entries in which are 0.

## III. Proposed Method

### A. Non-iterative Prediction-Correction Algorithm

In this section, we develop a non-iterative PC method to solve the time-varying optimization problem. Let $\boldsymbol{u}^*(t)$ be the optimal solution of problem (3). Notice that there are $p=2n+n$ time-varying inequality constraint functions included in (3b) and (3c), which can be compactly written as $f_i(\boldsymbol{u}, t) \le 0$ for $i \in \{1, 2, \ldots, p\}$, or $i \in [p]$. The following barrier function [5] is used to involve the inequality constraints into the objective function in problem (3):

$$\Phi(\boldsymbol{u},s,c,t) = f(\boldsymbol{u},t) - \frac{1}{c(t)}\sum_{i=1}^{p}\ln\left(s(t) - f_i(\boldsymbol{u},t)\right), \tag{4}$$

where the barrier parameter $c(t)$ is an increasing function satisfying $\lim_{t\to\infty} c(t) = \infty$ and the slack $s(t)$ is a decreasing function satisfying $\lim_{t\to\infty} s(t) = 0$, which is introduced to ensure that $s(t) > f_i(\boldsymbol{u}, t)$, $i \in [p]$ holds for all times $t \ge 0$. In particular, choosing $s(0) \ge \max_{i\in[p]}\{f_i(\boldsymbol{u}(0), 0)\}$ is sufficient to make this case true, as verified in [14]. Then, the barrier function takes the value 0 when the inequality constraints are satisfied and $+\infty$ in the opposite case.

Observe that the optimal solution $\hat{\boldsymbol{u}}^*(t)$ to (4) should satisfy the first-order optimality condition $\nabla_{\boldsymbol{u}}\Phi(\hat{\boldsymbol{u}}^*(t), s(t), c(t), t) = 0$, $\forall t \ge 0$, and thus its derivative should also satisfy formula (5):

$$\dot{\nabla}_{\boldsymbol{u}}\Phi = \nabla_{\boldsymbol{uu}}\Phi\dot{\hat{\boldsymbol{u}}}^*(t) + \nabla_{\boldsymbol{u}s}\Phi\dot{s} + \nabla_{\boldsymbol{u}c}\Phi\dot{c} + \nabla_{\boldsymbol{u}t}\Phi = 0, \tag{5}$$

where $\dot{\nabla}_{\boldsymbol{u}}\Phi$ represents the total derivative of $\nabla_{\boldsymbol{u}}\Phi$ to $t$, while $\nabla_{\boldsymbol{u}s}\Phi$, $\nabla_{\boldsymbol{u}c}\Phi$, and $\nabla_{\boldsymbol{u}t}\Phi$ denote the partial derivative of $\nabla_{\boldsymbol{u}}\Phi$ to $s$, $c$, and $t$, respectively. Solving (5) for $\hat{\boldsymbol{u}}^*(t)$ we obtain

$$\dot{\hat{\boldsymbol{u}}}^*(t) = -\nabla_{\boldsymbol{uu}}^{-1}\Phi\left(\nabla_{\boldsymbol{u}s}\Phi\dot{s} + \nabla_{\boldsymbol{u}c}\Phi\dot{c} + \nabla_{\boldsymbol{u}t}\Phi\right). \tag{6}$$

Equation (6) is called the "prediction" term, which is used to predict how the optimal solution changes over time by considering the time-varying characteristics of problem parameters. However, if we cannot obtain the initial optimal solution $\hat{\boldsymbol{u}}^*(0)$ or any optimal solution $\hat{\boldsymbol{u}}^*(t_0)$ for some $t_0 \ge 0$, then (6) could not be simply relied on to track the evolution of $\hat{\boldsymbol{u}}^*(t)$.

To push the solution to the optimum, Newton's method is implemented on $\Phi(\boldsymbol{u},s,c,t)$ to rapidly converge on its minimizer $\hat{\boldsymbol{u}}^*(t)$, the continuous-time version of which yields the following correction term:

$$\dot{\boldsymbol{u}}(t) = -\nabla_{\boldsymbol{uu}}^{-1}\Phi\left(\boldsymbol{u}(t), s(t), c(t), t\right)\nabla_{\boldsymbol{u}}\Phi\left(\boldsymbol{u}(t), s(t), c(t), t\right). \tag{7}$$

By combining (6) and (7), a complete prediction-correction dynamic system can be obtained:

$$\dot{\boldsymbol{u}}(t) = -\nabla_{\boldsymbol{uu}}^{-1}\Phi\left[\Lambda\nabla_{\boldsymbol{u}}\Phi + \nabla_{\boldsymbol{u}s}\Phi\dot{s} + \nabla_{\boldsymbol{u}c}\Phi\dot{c} + \nabla_{\boldsymbol{u}t}\Phi\right], \tag{8}$$

where $\Lambda = \alpha\boldsymbol{I}$ for some $\alpha > 0$, and $\boldsymbol{I}$ denotes an identity matrix.

Notice that in the dynamic system (8), the calculation of control increment $\dot{\boldsymbol{u}}$ does not require iteration, but can be obtained directly according to the right-side expression of the equal sign. Another thing worth mentioning is that the dynamical system (8) includes the prediction term $\nabla_{\boldsymbol{ut}}\Phi$, whose computation involves terms of DERs' available power and uncontrollable load drift over time. In online applications, perhaps only limited information about these terms is available. Here, assume the knowledge of the time-varying power within 1 s. Then Hermite polynomials can be used to approximate the continuous-time trajectory of the data sets with the desired level of accuracy [27].

Specifically for problem (3), the parameters in dynamic system (8) can be expressed as:

$$\nabla_{\boldsymbol{u}}\Phi = \boldsymbol{K}\boldsymbol{u} + \boldsymbol{d} + \gamma \boldsymbol{D}^T\left(\boldsymbol{V} - \boldsymbol{V}_{nom}\right) - \left(\boldsymbol{K}_1 + \boldsymbol{K}_2 + \boldsymbol{K}_3 + \boldsymbol{K}_4\right)/c, \tag{9}$$

$$\nabla_{\boldsymbol{uu}}\Phi = \boldsymbol{K} + \gamma \boldsymbol{D}^T\boldsymbol{D} - \left(\tilde{\boldsymbol{K}}_1 + \tilde{\boldsymbol{K}}_2 + \tilde{\boldsymbol{K}}_3 + \tilde{\boldsymbol{K}}_4\right)/c, \tag{10}$$

$$\nabla_{\boldsymbol{us}}\Phi = -\left(\boldsymbol{K}_{1s} + \boldsymbol{K}_{2s} + \boldsymbol{K}_{3s} + \boldsymbol{K}_{4s}\right)/c, \tag{11}$$

$$\nabla_{\boldsymbol{uc}}\Phi = \left(\boldsymbol{K}_1 + \boldsymbol{K}_2 + \boldsymbol{K}_3 + \boldsymbol{K}_4\right)/c^2, \tag{12}$$

$$\begin{aligned}\nabla_{\boldsymbol{ut}}\Phi &= \nabla_t \boldsymbol{d} + \boldsymbol{D}^T\boldsymbol{D}\nabla_t \boldsymbol{W} \\ &\quad -\left(\nabla_t \boldsymbol{K}_1 + \nabla_t \boldsymbol{K}_2 + \nabla_t \boldsymbol{K}_3 + \nabla_t \boldsymbol{K}_4\right)/c\end{aligned}, \tag{13}$$

Let $h_i^{'} = s - u_i^{\min} + u_i$, $h_i^{''} = s - u_i + u_i^{\max}$, $g_i^{'} = s - V_i^{\min} + V_i$, $g_i^{''} = s - V_i + V_i^{\max}$, and $\boldsymbol{e}_i$ be a basis vector whose $i$-th element is 1 and the other elements are 0. Then the specific parameters in (9)–(13) can be expressed as

$$\begin{cases} \boldsymbol{K}_1 = \left[1/h_i^{'}\right]_{2n\times 1}, \boldsymbol{K}_2 = \left[-1/h_i^{''}\right]_{2n\times 1}, \\ \boldsymbol{K}_{1s} = \left[-1/(h_i^{'})^2\right]_{2n\times 1}, \boldsymbol{K}_{2s} = \left[1/(h_i^{''})^2\right]_{2n\times 1}, \\ \tilde{\boldsymbol{K}}_1 = diag\left(-1/(h_i^{'})^2\right), \tilde{\boldsymbol{K}}_2 = diag\left(-1/(h_i^{''})^2\right) \\ \nabla_t \boldsymbol{K}_1 = \left[\dot{u}_i^{\min}/(h_i^{'})^2\right]_{2n\times 1}, \nabla_t \boldsymbol{K}_2 = \left[\dot{u}_i^{\max}/(h_i^{''})^2\right]_{2n\times 1} \end{cases}, i \in [2n],$$

$$\begin{cases} \boldsymbol{K}_3 = \sum_{i=1}^{n} 1/g_i^{'}\boldsymbol{D}^T\boldsymbol{e}_i, \boldsymbol{K}_4 = \sum_{i=1}^{n} -1/g_i^{''}\boldsymbol{D}^T\boldsymbol{e}_i, \\ \boldsymbol{K}_{3s} = \sum_{i=1}^{n} -1/\left(g_i^{'}\right)^2\boldsymbol{D}^T\boldsymbol{e}_i, \boldsymbol{K}_{4s} = \sum_{i=1}^{n} 1/\left(g_i^{''}\right)^2\boldsymbol{D}^T\boldsymbol{e}_i, \\ \tilde{\boldsymbol{K}}_3 = \sum_{i=1}^{n} -1/\left(g_i^{'}\right)^2\left(\boldsymbol{e}_i^T\boldsymbol{D}\right)^T\left(\boldsymbol{e}_i^T\boldsymbol{D}\right), \\ \tilde{\boldsymbol{K}}_4 = \sum_{i=1}^{n} -1/\left(g_i^{''}\right)^2\left(\boldsymbol{e}_i^T\boldsymbol{D}\right)^T\left(\boldsymbol{e}_i^T\boldsymbol{D}\right), \\ \nabla_t \boldsymbol{K}_3 = \sum_{i=1}^{n} -1/\left(g_i^{'}\right)^2\left(\boldsymbol{e}_i^T\boldsymbol{D}\right)^T\left(\boldsymbol{e}_i^T\boldsymbol{D}\right)\nabla_t \boldsymbol{W}, \\ \nabla_t \boldsymbol{K}_3 = \sum_{i=1}^{n} -1/\left(g_i^{''}\right)^2\left(\boldsymbol{e}_i^T\boldsymbol{D}\right)^T\left(\boldsymbol{e}_i^T\boldsymbol{D}\right)\nabla_t \boldsymbol{W}, \end{cases}, i \in [n],$$

$$\nabla_t \boldsymbol{d} = \left[-2C_{pi}dP_i^{tar}/dt, i \in N \quad \boldsymbol{0}\right]^T,$$

$$\nabla_t \boldsymbol{W} = \left[\nabla_t \boldsymbol{P}_L^{\,T} \quad \nabla_t \boldsymbol{Q}_L^{\,T}\right]^T.$$

### *B. Estimation of Voltage Sensitivity Parameters*

In this section, the nonlinear relationship between $\boldsymbol{V}$, $\boldsymbol{P}$, and $\boldsymbol{Q}$ is approximated by the linear model (2), and the time-varying sensitivity parameters $\boldsymbol{A}$ and $\boldsymbol{B}$ will be estimated by measurements.

Let $\tilde{\boldsymbol{M}} \in \mathbb{R}^{(n+1)\times L}$ represent the node-to-branch incidence matrix of the distribution network, and $\boldsymbol{M}$ be a matrix obtained from $\tilde{\boldsymbol{M}}$ by removing the row associated with the PCC. Specially, in balanced radial distribution networks, voltage sensitivity matrices can be expressed as:

$$\boldsymbol{A} = \left(\boldsymbol{M}^{-1}\right)^T diag\left(\tilde{\boldsymbol{r}}\right)\boldsymbol{M}^{-1}, \tag{14a}$$

$$\boldsymbol{B} = \left(\boldsymbol{M}^{-1}\right)^T diag\left(\tilde{\boldsymbol{x}}\right)\boldsymbol{M}^{-1}, \tag{14b}$$

where $\tilde{\boldsymbol{r}}$ and $\tilde{\boldsymbol{x}}$ $\in \mathbb{R}^{L\times 1}$ whose elements represent the correlation coefficients between two connected nodes, and they are equivalent to line resistance and line reactance respectively under no load conditions.

Assuming the network topology configurations $\boldsymbol{M}$ (which would not change frequently within a short period) and distribution line resistance-to-reactance ratios $\varsigma_l$ are available, then $\tilde{\boldsymbol{r}}$ and $\tilde{\boldsymbol{x}}$ can be identified using only a few measurements in the design. At time $t_{k+1}$, assume the knowledge of voltage magnitude measurements $\boldsymbol{V}[k]$, and power injections $\boldsymbol{P}[k]$ and $\boldsymbol{Q}[k]$, where $k \in \kappa = \{t_{k-m}, \cdots, t_k\}$. Let $\boldsymbol{v}[k] = \boldsymbol{V}[k] - \boldsymbol{1}_n$, and $\hat{\boldsymbol{r}}$ and $\hat{\boldsymbol{x}}$ represent the estimates of $\tilde{\boldsymbol{r}}$ and $\tilde{\boldsymbol{x}}$, respectively. The parameter estimation problem can be constructed as follows [26]:

$$\left(\hat{\boldsymbol{r}}, \hat{\boldsymbol{x}}\right) = \arg\min_{\tilde{\boldsymbol{r}}, \tilde{\boldsymbol{x}}} \sum_{k\in\kappa} \eta^{t_k - k}\left\|\boldsymbol{A}\boldsymbol{P}[k] + \boldsymbol{B}\boldsymbol{Q}[k] - \boldsymbol{v}[k]\right\|^2, \tag{15a}$$

$$s.t \quad (14a), (14b), \tag{15b}$$

where $\|\cdot\|^2$ denotes the $L_2$-norm, and $\eta \in (0,1]$ is a forgetting factor.

Let $\boldsymbol{\chi}_l[k] = \eta^{(t_k - k)/2}(\varsigma_l \boldsymbol{P}[k] + \boldsymbol{Q}[k])$, $\boldsymbol{\Upsilon}_l = (\boldsymbol{M}^{-1})^T\boldsymbol{e}_l\boldsymbol{e}_l^T\boldsymbol{M}^{-1}$, and define

$$\mathbf{Z}[t_k] = \begin{bmatrix} \boldsymbol{\Upsilon}_1\boldsymbol{\chi}_1[t_{k-m}] & \cdots & \boldsymbol{\Upsilon}_L\boldsymbol{\chi}_L[t_{k-m}] \\ \vdots & \ddots & \vdots \\ \boldsymbol{\Upsilon}_1\boldsymbol{\chi}_1[t_k] & \cdots & \boldsymbol{\Upsilon}_L\boldsymbol{\chi}_L[t_k] \end{bmatrix}, \tag{16}$$

and

$$\varphi[t_k] = \left[\eta^{\frac{m}{2}}\boldsymbol{v}[t_{k-m}]^T, \cdots, \eta^{\frac{0}{2}}\boldsymbol{v}[t_k]^T\right]^T, \tag{17}$$

Then (15) is equivalent to the following classical linear regression problem:

$$\min_{\hat{\boldsymbol{x}}}\left\|\mathbf{Z}[t_k]\hat{\boldsymbol{x}} - \varphi[t_k]\right\|^2, \tag{18}$$

whose closed-form solution is:

$$\hat{\boldsymbol{x}} = \mathbf{Z}[t_k]^{\dagger}\varphi[t_k], \tag{19}$$

where $\mathbf{Z}[t_k]^{\dagger} = (\mathbf{Z}[t_k]^T\mathbf{Z}[t_k])^{-1}\mathbf{Z}[t_k]^T$ is the pseudo-inverse of $\mathbf{Z}[t_k]$. Then the estimated vectors $\hat{\boldsymbol{A}}$ and $\hat{\boldsymbol{B}}$ can be calculated as:

$$\hat{\boldsymbol{A}} = \left(\boldsymbol{M}^{-1}\right)^T diag\left(diag(\varsigma)\mathbf{Z}[t_k]^{\dagger}\varphi[t_k]\right)\boldsymbol{M}^{-1}, \tag{20a}$$

$$\hat{\boldsymbol{B}} = \left(\boldsymbol{M}^{-1}\right)^T diag\left(\mathbf{Z}[t_k]^{\dagger}\varphi[t_k]\right)\boldsymbol{M}^{-1}. \tag{20b}$$

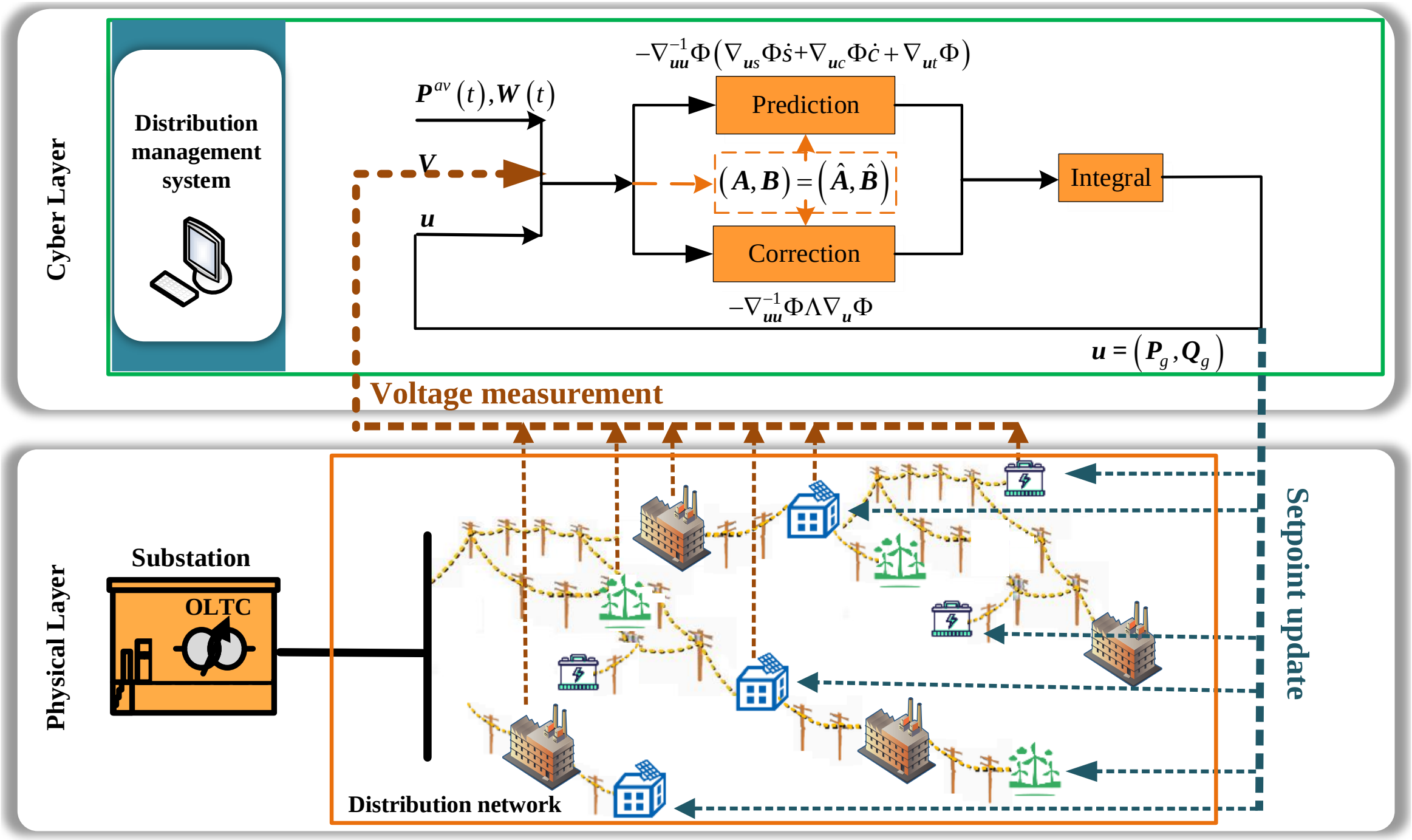

Fig. 1. Schematic diagram of the proposed non-iterative algorithm framework.

Notice that (20) is a closed-form formulation, only related to the known parameters (e.g., network topology parameters and power injections) and a few measurements. Therefore, the proposed algorithm can be extended to application in the absence of an accurate network without incurring extra computational burden.

### C. Proposed Algorithm Framework

The schematic diagram of the presented algorithm framework is highlighted in Fig. 1. A dynamical system consisting of "prediction" and "correction" terms for distribution grids is developed to identify the time-varying optimal power setpoints of DERs and network voltages, which automatically adapt to system disturbance, such as renewable energy sources, uncontrollable loads, and system model parameters.

In actual execution, the output power increments of DERs are obtained based on the predictive power parameters in the ultra-short time and their gradients with respect to time, as well as the voltage measurements. Setpoints of DERs are obtained by integrating the increments over time, which will affect the system voltage through network constraints. After obtaining the new power parameters and network voltages, the new increments of DER setpoints can be calculated. That is to say, in each interval, the proposed algorithm only needs to perform a single one-step calculation instead of performing multiple iterations while guaranteeing a good tracking performance.

## IV. Simulation

The performance of the proposed method was evaluated using several numerical experiments on a modified IEEE 33-bus distribution network [28]. The first experiment under normal time-varying conditions was carried out to verify the tracking performance of the proposed non-iterative algorithm compared with the optimal trajectories intuitively, together with comparisons with iterative algorithms and fixed sensitivity methods. The second experiment with one PV halting and resuming operation suddenly was presented to demonstrate the robustness performance of our approach under abnormal scenarios. The third experiment tested a network reconfiguration scenario to show the benefits of the method's adaptivity to changes in the system model. The dynamic optimization system illustrated in Fig. 1 was established in a MATLAB/Simulink platform. We numerically solve the dynamical system in (8) with step size $\tau = 0.02$ s. The computer was equipped with an Intel(R) Core(TM) processor i7-4800MQ with 16 GB of RAM.

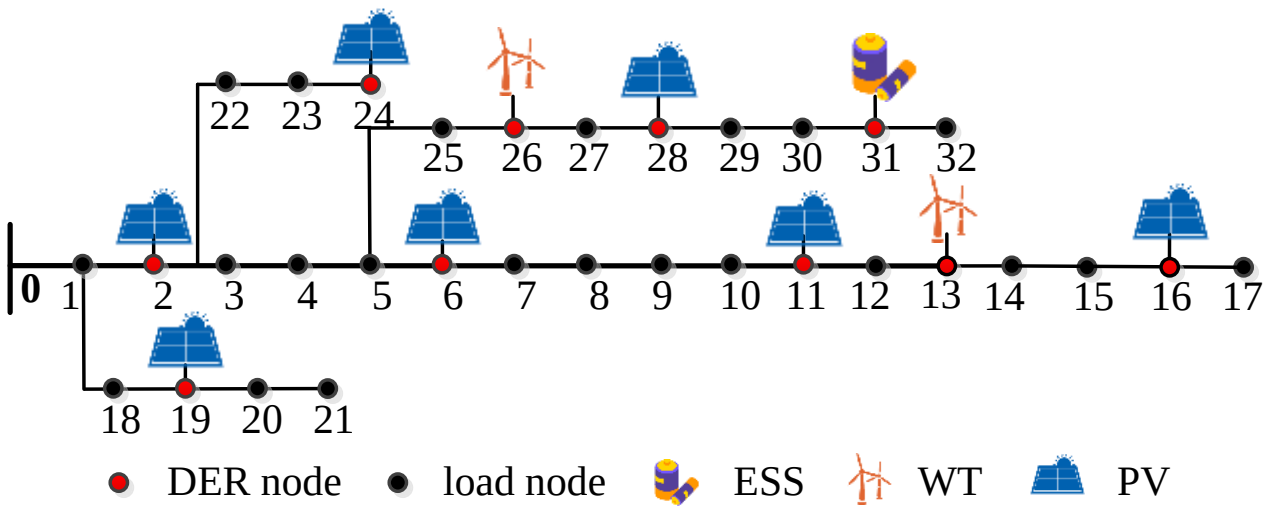

Fig. 2. IEEE 33-bus distribution network topology.

Fig. 2 shows the topology of the IEEE 33-bus distribution network, where PVs were located at nodes {2, 6, 11, 16, 19, 24, 28}, WTs were located at nodes {13, 26}, and an ESS was located at node {31}. The available real power of DERs [29] and real power loads [30] are given in Fig. 3. Consider that all loads follow the daily real power shown in Fig. 3 with constant power factors. The parameters of the ESS are $P_{\max}^{ch} = P_{\max}^{dis} = 200$ kW, $\eta_c = \eta_d = 0.9$, $W_{\min} = 80$ kWh, and $W_{\max} = 320$ kWh. Let voltage limits $V_i^{\min} = 0.95$ p.u. and $V_i^{\max} = 1.05$ p.u., and the nominal value $\mathbf{V}_{nom} = 1$ p.u. Other parameters for the proposed optimization problem are as follows: $C_p = 3$,

$C_q = 1$, $\gamma = 1$, $\alpha = 100$, $m = 1$, $\eta = 1$, $s(t) = 2e^{-10t}$, $c(t) = e^{10t}$, and $\cos\theta = 0.85$.

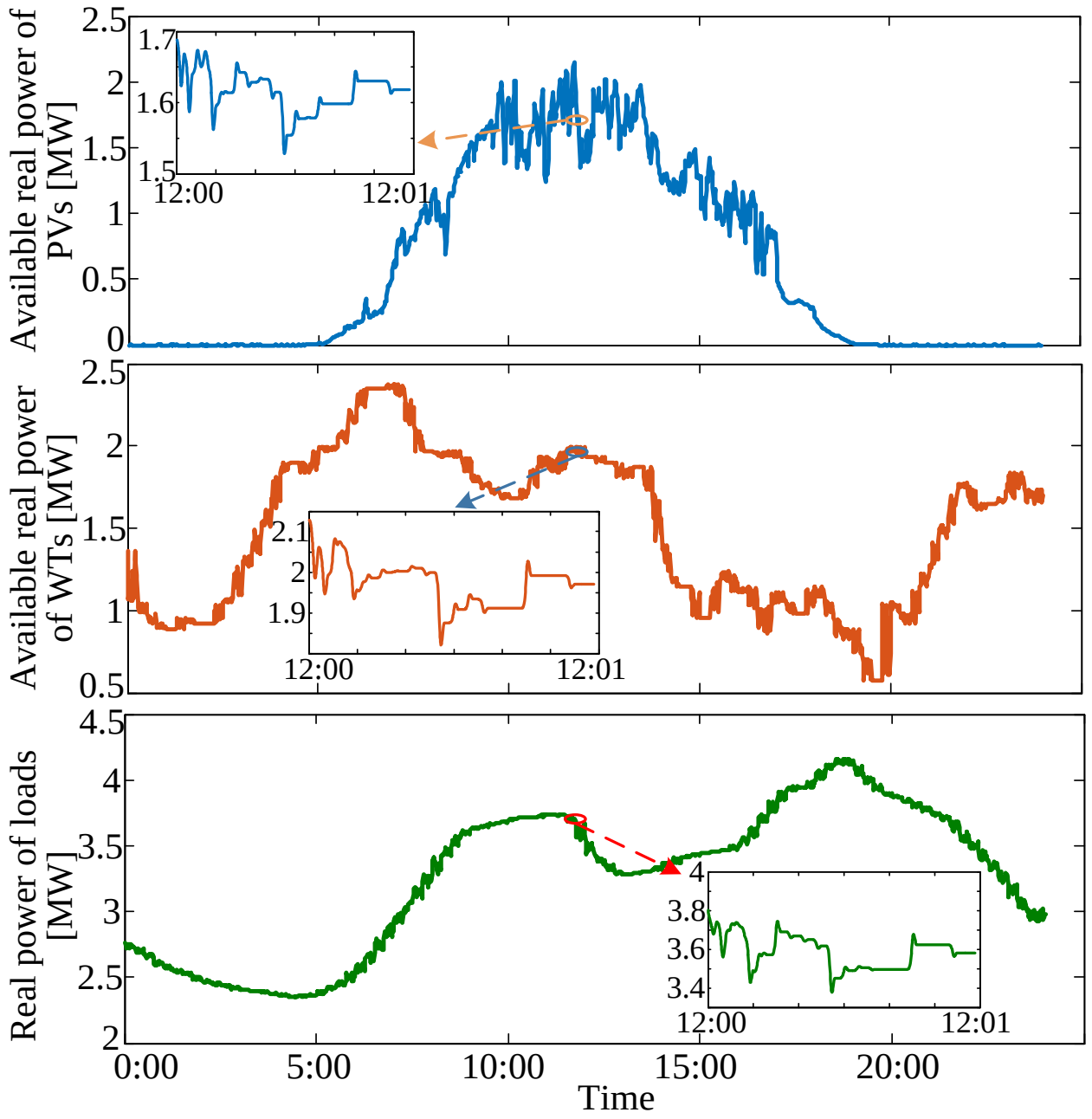

Fig. 3. Trajectories of DERs available power and load.

### A. Dynamic Trajectories under Normal Time-Varying Conditions

First, a test under normal time-varying conditions was carried out at 12:00–12:01 p.m. A Gaussian distribution model $\varepsilon \sim N(0,\sigma^2)$ with $\sigma^2$=0.001 was added to the rated line reactance parameters specified in [28] to simulate uncertain feeder parameters.

*1) Results comparison with exact optimal solution:* To verify the efficiency and tracking performance of the proposed algorithm, exact optimal trajectories obtained by sampling the problem every 0.02 s and solved using the solver of the "yalmip" toolbox in MATLAB are used for comparison. Notice that here we assume that the solver has sufficient time to obtain the optimal solution within each sampling interval, whose real computing time will be analyzed later. The trajectories of power outputs of DERs and voltage magnitudes of some example nodes are depicted in Fig. 4. The dashed lines represent the track trajectories calculated by the proposed algorithm, while the solid lines represent the exact optimal trajectories.

As shown in Fig. 4, starting from the initial values, trajectories of the control variables (i.e., the real and reactive power setpoints of DERs) and voltage magnitudes calculated by the proposed method gradually approached the optimal solution, tracked to the exact optimal trajectories after about 1.5 s, and stayed on that optimum in the subsequent time-varying process. This means that as long as we start the algorithm, we will be able to identify the exact optimum within a short period of time (e.g., 1.5 s) compared to the entire tracking time, which can be set according to the specific situation and is set to be 60 s in this case. The algorithm will remain on the optimal trajectory regardless of whether the parameters change over time.

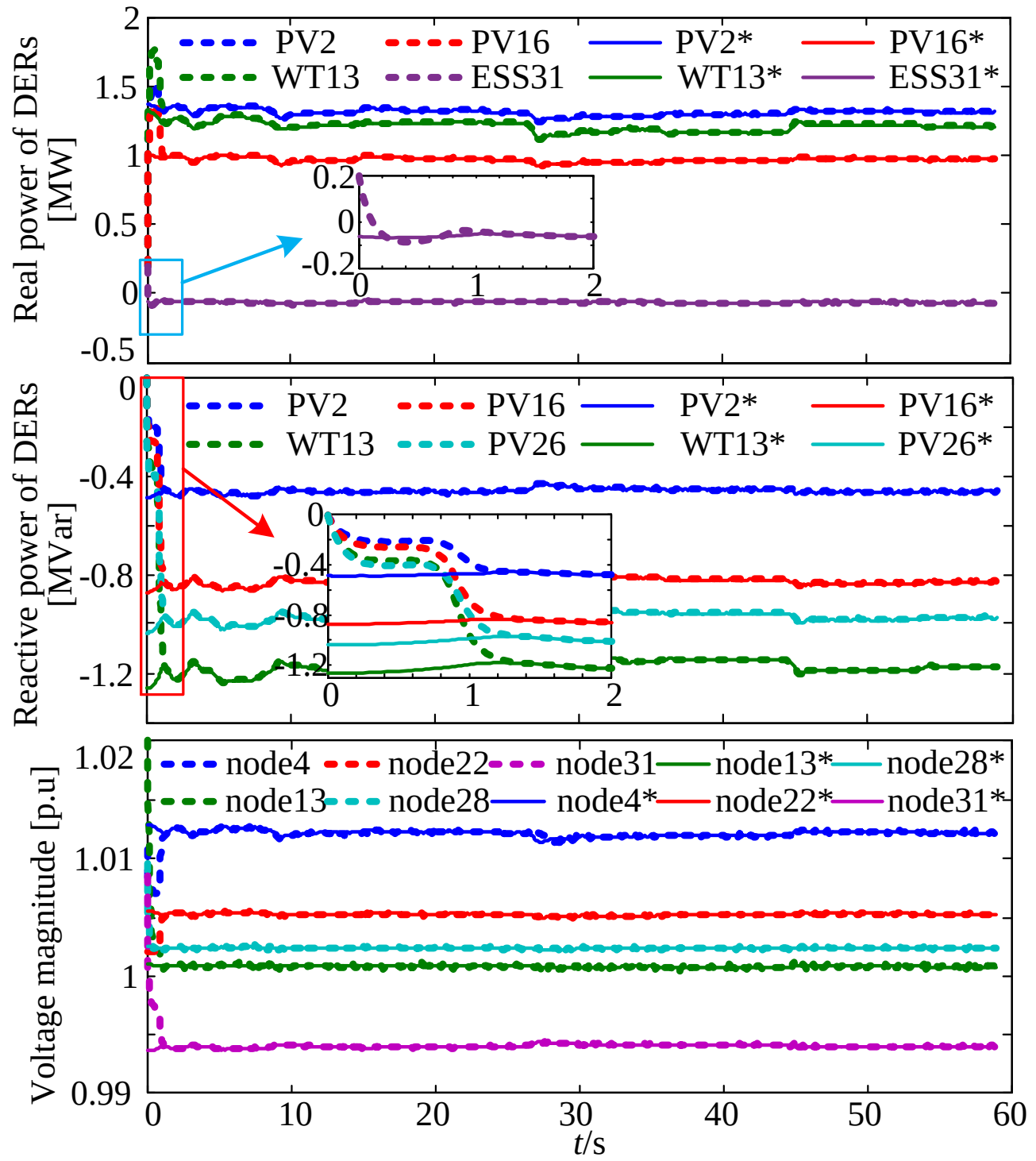

Fig. 4. Dynamic optimal trajectories of setpoints of DERs and voltage magnitudes under normal time-varying conditions.

To demonstrate the accuracy of our proposed method, we plot the $L_2$-norm of errors of control variables and the objective value versus $t$ in Fig. 5. It can be found that the tracking errors drop rapidly and remain within a small range, such as $10^{-3}$–$10^{-4}$ for control variables and $10^{-4}$–$10^{-5}$ for the objective value.

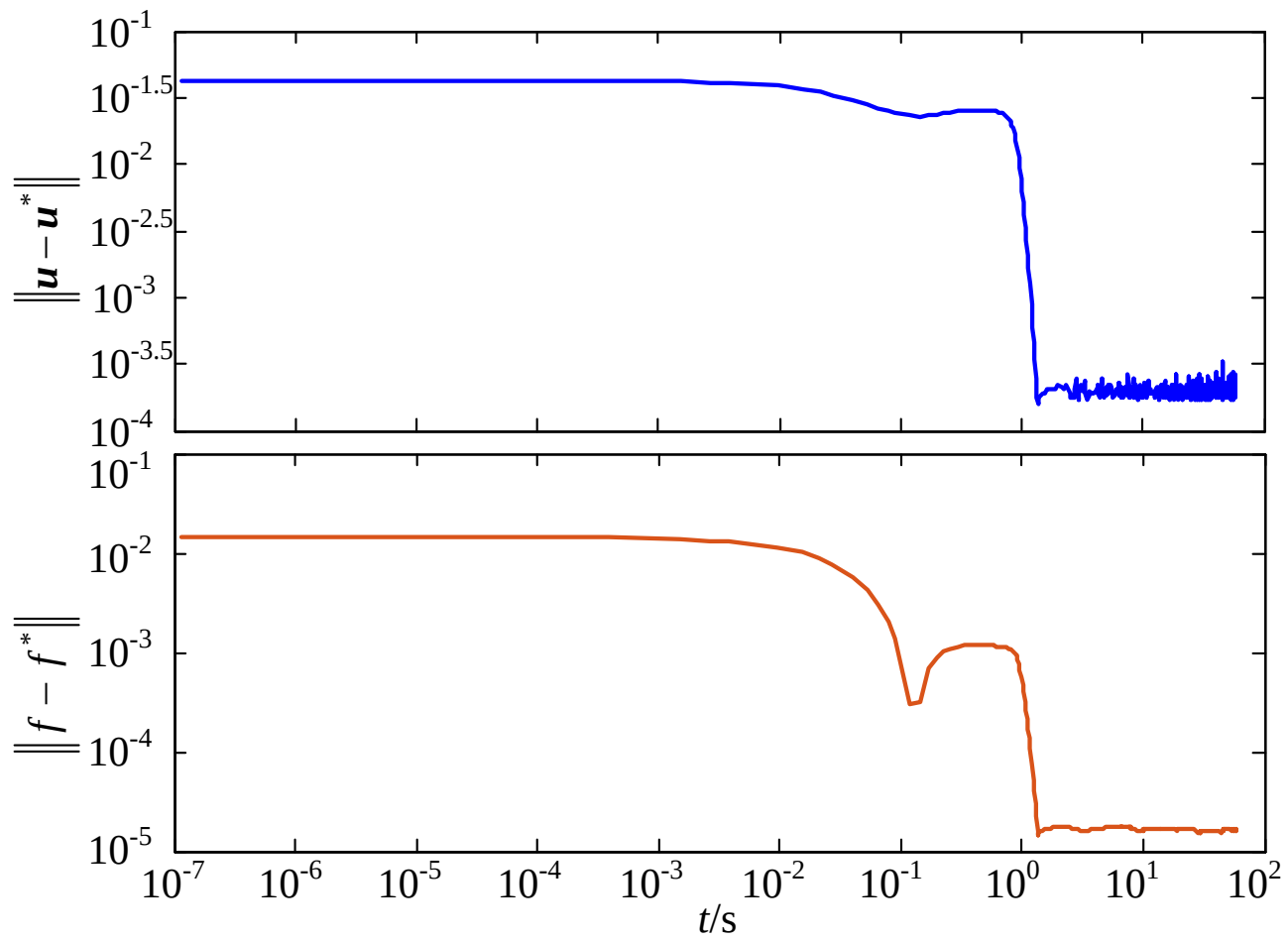

Fig. 5. Error trajectories of control variables and objective value.

*2) Results comparison with iterative algorithms:* To evaluate the advantages of the proposed non-iterative algorithm, we compared the proposed algorithm with some iterative algorithms, including the primal-dual algorithm in [12] (without a "prediction" step), and the discrete-time PC algorithm in [19]. For iterative algorithms, here, we sample the time-varying optimization problem every 1 s, and iterate multiple times in the interval until convergence. The exact optimal trajectories obtained by the same way as in 1) were used to be compared. The results of real power setpoint of WT13 are shown in Fig. 6.

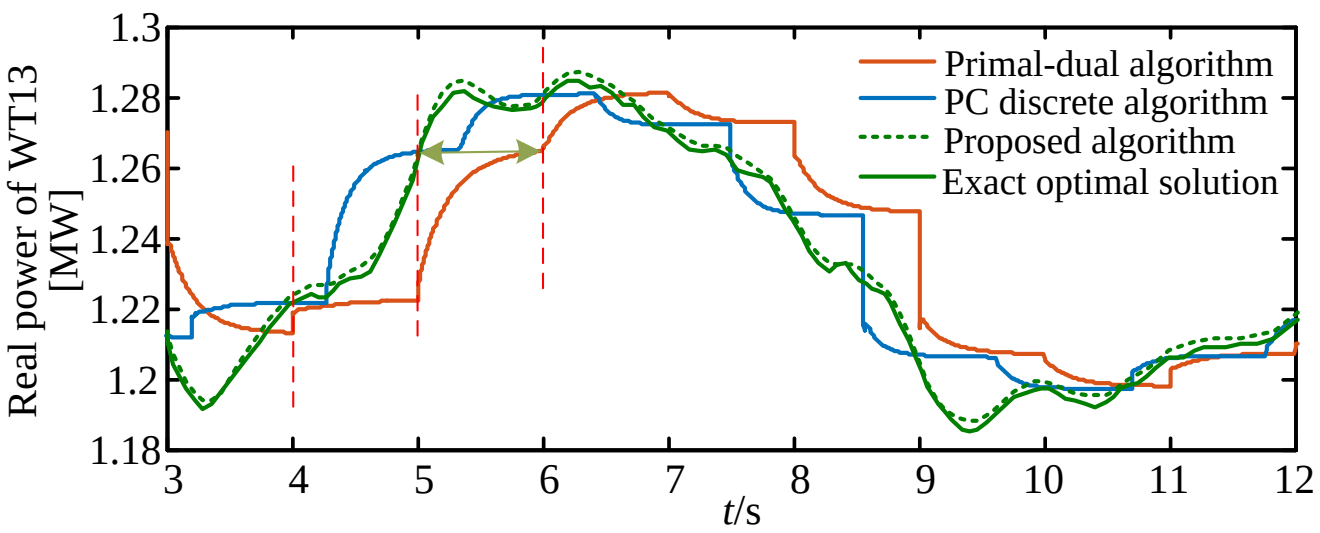

Fig. 6. Result comparison of non-iterative and iterative algorithms.

In terms of tracking performance, it can be seen that the trajectory obtained by the proposed algorithm (the green dotted line) can track the exact optimal trajectory (the green solid line) well—the tracking error is so small that it is almost negligible. The result obtained by the primal-dual algorithm (the orange solid line) always lags behind the exact optimal trajectory, since it tends to converge toward the optimal solution of the sampled problem, while the actual optimum is drifting with time, leading to a steady-state deviation. The discrete-time PC algorithm incorporates the prediction of time-varying parameters so that the result (the blue solid line) has little deviation at the sampling points. However, it still cannot fully capture the dynamics in the interval, leading to a large error within the sampling interval.

We compare the computation time of each algorithm in Table I.

TABLE I
COMPUTATION TIME COMPARISON OF DIFFERENT ALGORITHMS.

| Algorithms | Primal-dual | Discrete-time PC | Proposed |
|---|---|---|---|
| Time of each iteration (s) | $3.95\times10^{-4}$ | $3.67\times10^{-4}$ | 0.0117 |
| Time to get the optimal solution (s) | 0.79 (with tracking deviation) | 0.661 | 0.0117 |

As can be seen, it took the primal-dual and discrete-time PC algorithms 0.79 s and 0.661 s, respectively, to produce solutions with tracking error. It took 0.0117 s for the proposed algorithm to iterate one step, while the optimal solution can be obtained in each iteration. This means that it can be applied in real time at intervals of 0.02 s to match the DER's regulation time. According to our tests, the real computing time for the solver to solve the sampling problem is 0.419 s, which means that it is difficult for the solver to obtain the optimal solution in the application interval of 0.02 s in practice. Here, we obtain the exact optimal solution just for the sake of comparison, regardless of the time cost.

*3) Results comparison with fixed sensitivity method:* To evaluate the advantages of the sensitivity estimation method, we compared the tracking errors in Fig. 7 between the proposed algorithm with estimated sensitivities and the simplified version of the proposed algorithm with fixed sensitivities (calculated from the rate line parameters [32]). As shown, the errors of both the control variables and the objective value of the fixed sensitivity method are larger and more oscillating than those of the proposed estimated sensitivity method.

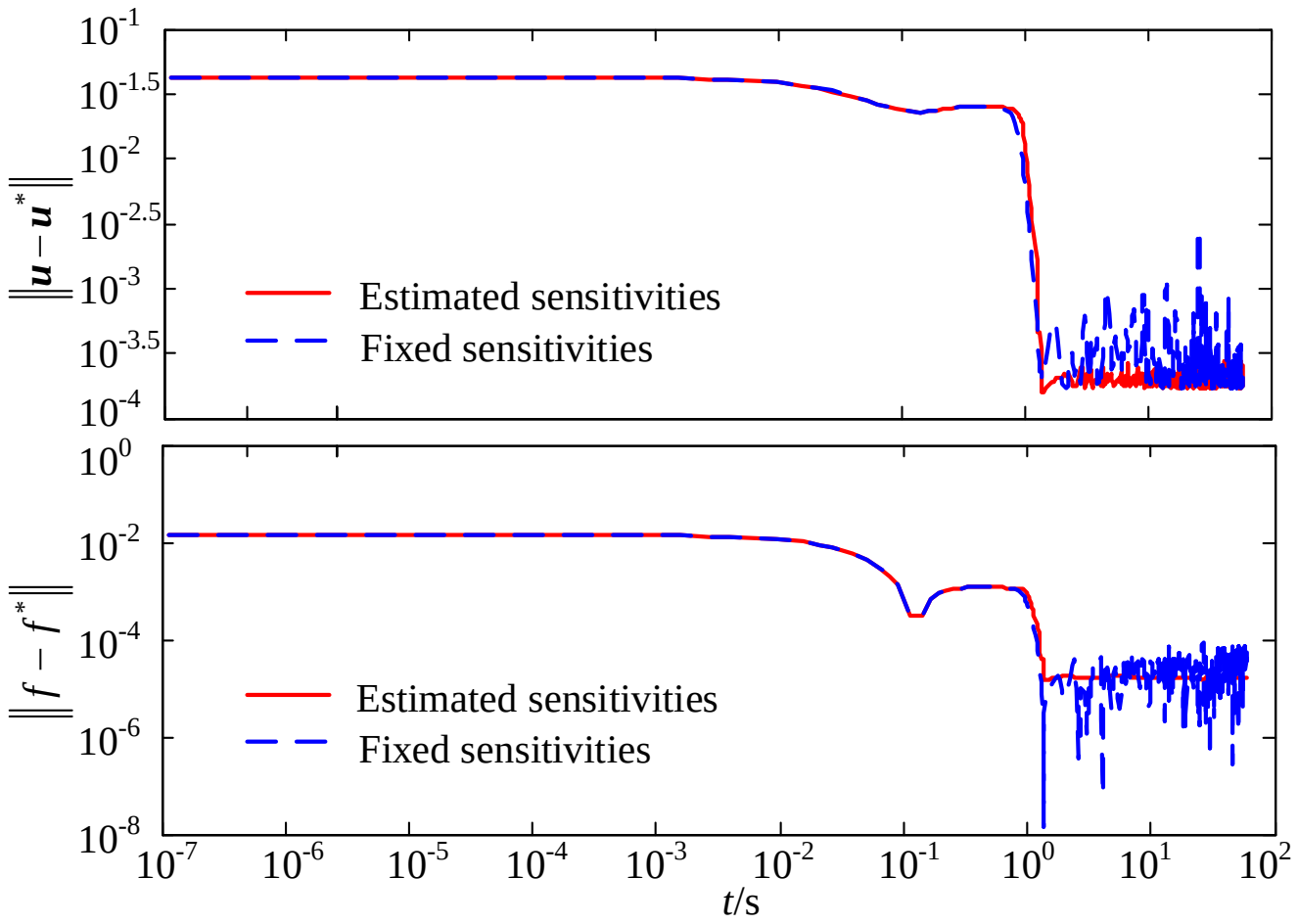

Fig. 7. Result comparison of methods with estimated and fixed sensitivities.

## *B. Dynamic Trajectories when One PV Is Abnormal*

To test the robustness performance of the proposed method in an abnormal scenario, we assumed that PV28 halted operation at time $T$ = 12:00:25 and resumed operation after 10 s, and showed the test results in Fig. 8.

As shown in Fig. 8, when PV28 suddenly halts operation, the other DERs begin to react and reach new optima at short notice. Especially, the ESS deployed at node 31 was changed from the original charging state to the discharge state to supplement the power deficiency of the system (shown in the upper figure of Fig. 8). When PV28 resumed operation, ESS31 returned to the charging state.

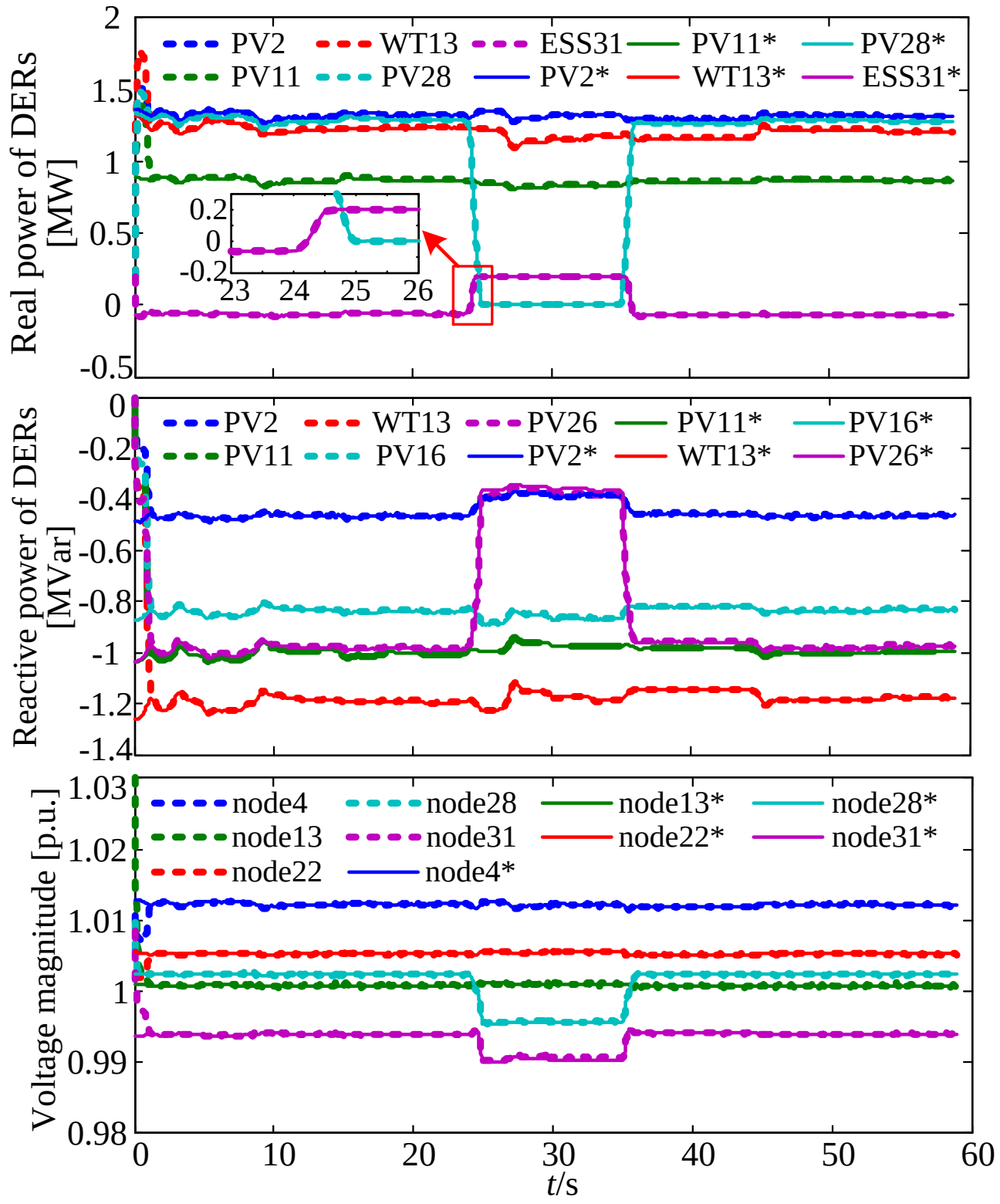

Fig. 8. Dynamic optimal trajectories of setpoints of DERs and voltage magnitudes when one PV is abnormal.

### C. Dynamic Trajectories when Network Reconfiguration

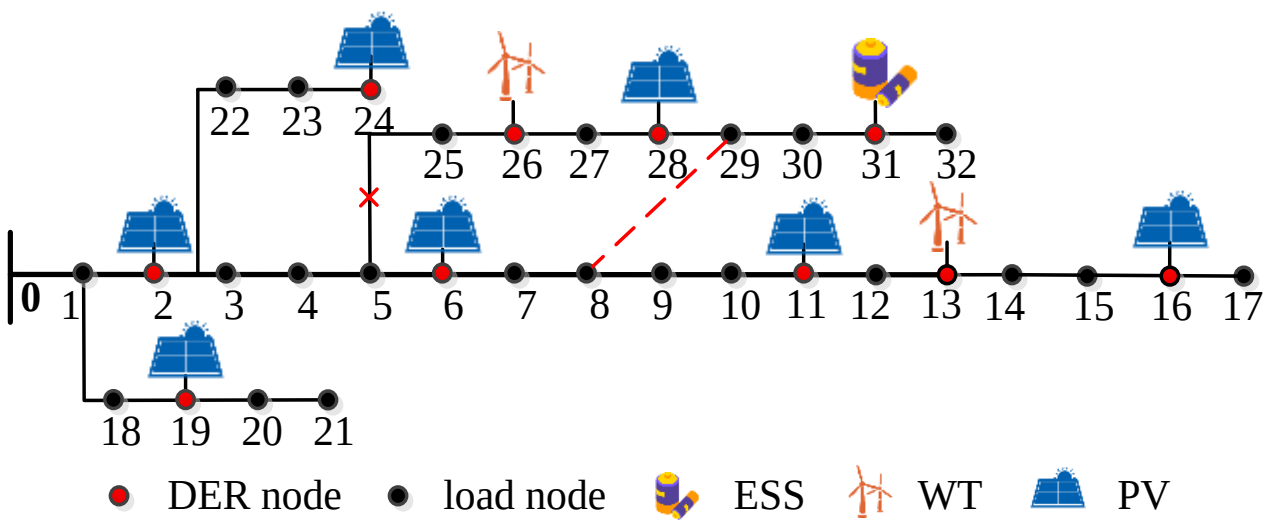

Fig. 9. IEEE 33-bus distribution network reconfiguration topology.

A network reconfiguration scenario was tested at time $T$ = 12:00:30 where bus 8 was connected to bus 29, and the line between bus 5 and bus 25 was disconnected, as shown in Fig. 9. Fig. 10 shows results of power outputs of DERs and voltage magnitudes, from which we can see that the trajectories obtained by the proposed algorithm can return quickly to the optimal trajectories when the network topology changes, taking about 0.5 s.

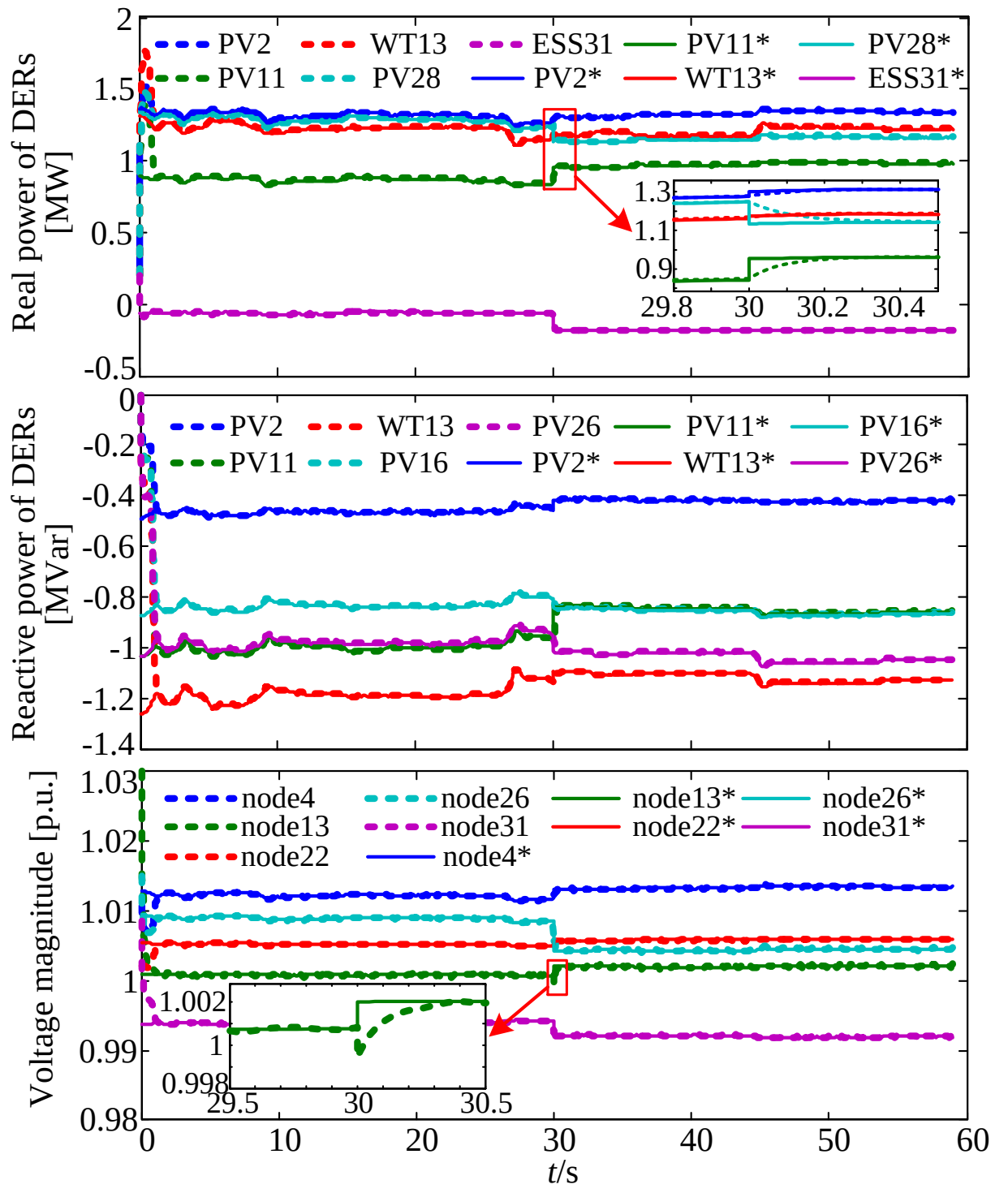

Fig. 10. Dynamic optimal trajectories of setpoints of DERs and voltage magnitudes when network reconfiguration.

## V. Conclusion

In the present study, a non-iterative algorithm was designed to track the time-varying optimal power setpoints of DERs and network voltages for distribution grids with an asymptotically vanishing error. Considering the fast-changing nature of renewable energy resources and uncontrolled loads, we derive a prediction term to identify the change of the optimum; in addition, a correction term is formed based on Newton's method to push the solution toward the optimum. The proposed algorithm can be applied online in the absence of an accurate network model. Simulation results demonstrate that the proposed method is applicable for fast-changing loads and power generations, and variable feeder parameters.